\documentclass[%
 reprint,
 amsmath,amssymb,
 aps,
]{revtex4-2}

\usepackage{graphicx}
\usepackage{dcolumn}
\usepackage{bm}

\usepackage{threeparttable} 
\usepackage[colorlinks = true, linkcolor = blue, urlcolor  = blue, citecolor = blue, anchorcolor = blue]{hyperref}
\usepackage{float}
\usepackage{multirow}
\usepackage{ulem}
\usepackage{comment}
\begin{document}

\preprint{APS/123-QED}

\title{Microquasar Remnants as Pevatrons Illuminating the Galactic Cosmic Ray Knee}

\author{Bing Theodore Zhang$^{1,2}$}
\email[]{zhangbing@ihep.ac.cn}

\affiliation{$^1$ Key Laboratory of Particle Astrophysics and Experimental Physics Division and Computing Center, Institute of High Energy Physics, Chinese Academy of Sciences, 100049 Beijing, China}
\affiliation{$^2$ TIANFU Cosmic Ray Research Center, Chengdu, Sichuan, China}

\author{Shiqi Yu}
\email{shiqi.yu@utah.edu}
\affiliation{Department of Physics and Astronomy, University of Utah, Salt Lake City, Utah, USA}

\date{\today}

\begin{abstract} Microquasars are primary candidates for Galactic PeVatrons, yet their collective contribution to the cosmic ray (CR) ``knee" remains poorly understood. We investigate this contribution by simulating anisotropic diffusive propagation through the Galactic magnetic field. Our results demonstrate that spatial alignment and magnetic connectivity between source locations and the solar neighborhood govern the local flux: sources aligned with local magnetic field lines yield pronounced flux enhancements, whereas magnetically disconnected locations are strongly suppressed. 
We find that the characteristic proton bump near PeV is robust across random realizations of the Galactic microquasar remnant population, with as few as four nearby remnants accounting for 50\% of the observed PeV flux. Our findings suggest that the integrated history of microquasar remnants, governed by source temporal and spatial distribution and magnetic transport, naturally populates the observed CR ``knee''.
\end{abstract}

\pacs{Valid PACS appear here}
\maketitle


\noindent\textit{Introduction}\label{sec:intro}--
Recent observations by the LHAASO collaboration have identified a prominent ``bump" in the cosmic ray (CR) proton spectrum from 100 TeV to 30 PeV~\cite{LHAASO:2025byy}, which is coincident with the position of the ``knee" of the all-particle spectrum~\cite{LHAASO:2024knt}. This spectral feature necessitates a distinct population of Galactic PeVatrons~\cite{LHAASO:2025byy, LHAASO:2025mlf}, suggesting that the observed CR is shaped by specific accelerator classes, such as microquasars~\cite[e.g.,][]{Zhang:2025tew, Yue:2026gdt, Wang:2025yqy, Kaci:2025gyb,Fang:2026ydz,LHAASO:2024psv,LHAASO:2025ysm, Abaroa:2025ege}. 
Our previous work has introduced the physical foundations for microquasars as primary PeVatron candidates~\cite{Aharonian:2026tzf} and demonstrated that shear acceleration within jet-cocoon systems~\cite{Zhang:2025tew} can theoretically fuse the CRs to reproduce the observed PeV proton bump. 
However, the Galactic magnetic field (GMF) channels CRs preferentially along field lines, giving rise to anisotropic diffusion that modifies the observed CR spectra non-negligibly~\cite{Merten:2017mgk, AL-Zetoun:2025fxg,Neronov:2024ycp}. 
This magnetic channeling implies that two equidistant sources may yield vastly different observed fluxes depending on their connectivity to the solar neighborhood.

In this paper, we extend our previously developed CR model~\cite{Zhang:2025tew} by incorporating detailed propagation through the GMF. Specifically, we evaluate the cumulative contribution of CRs released by microquasars in the past, i.e., remnants, during active jet phases, which continue to diffuse through the interstellar medium (ISM) long after active injection has ceased. 

Our results demonstrate that the interplay between remnant source distribution and anisotropic transport in the GMF is essential to reproduce the observed data, with a hard injection spectrum favored to satisfy the required energy budget. By comparing our model against observed CR flux, we establish upper bounds on the CR loading factor ($\eta_{\rm CR}$). Our study provides a comprehensive assessment of the collective contribution from the microquasar population to the observed PeV CR proton spectrum. We note that while the halo boundary introduces a normalization degeneracy with $\eta_{\rm CR}$, a systematic exploration of transport models and parameters is beyond the scope of this work.

\noindent\textit{Methodology and Models}\label{sec:propa}--
We simulate CR propagation using \texttt{DiffusiveSDE} module in the \textsc{CRPropa 3.2} framework~\cite{CRPropa:2022ovg}. The GMF is represented by the JF12 model~\cite{Jansson:2012rt, 2019ApJ...877...76K}. The expected CR flux at Earth is estimated by recording particles entering a spherical volume of radius 200~pc centered at the solar neighborhood. This observing volume is chosen to be large enough to ensure statistical convergence of the stochastic particles while remaining small relative to Galactic scales, thereby providing a robust estimate of the local CR density. Since this work focuses on interpreting the time-integrated CR flux, we safely neglect uncertainties arising from jet variability~\cite{Fender:1998bb} and assume a constant CR injection rate. We further assume that CRs are injected into the ISM immediately after escaping from the sources and neglect the possible effects arising from jet duty cycles~\cite{2011A&A...528A..89B}.

For a source with an injection power $Q(E)$ [erg s$^{-1}$], the total injected CR energy is $W_{\rm cr}=Q\,\tau_{\rm dur}$, where $\tau_{\rm dur}$ is the active lifetime of the central engine and $L_{\rm cr}=\eta_{\rm cr}L_{\rm Edd}$ represents the total CR injection power expressed as a fraction ($\eta_{\rm CR}$) of the Eddington luminosity. The observed particle density $n(E, \mathbf{r}, \tau_{\rm age})$ at position $\mathbf{r}$ is obtained by integrating the source's injection history:
\begin{equation}\label{eq:n_particle}
n(E, \mathbf{r}, \tau_{\rm age}) = Q(E) \int_{\tau_{\min}}^{\tau_{\rm age}} \mathcal{P}(\mathbf{r}, E, \tau)  d\tau,\end{equation}
where $\mathcal{P}(\mathbf{r}, E, \tau)$ acts as the probability density function for a particle's position after a propagation time $\tau$.
The integration limits are defined by $\tau_{\min} = \max(0, \tau_{\rm age} - \tau_{\rm dur})$, where $\tau_{\rm age}$ represents the time elapsed since the jet onsets and $\tau_{\rm dur}$ is the engine's lifetime.
The observed flux is then $J(E) = (c/4\pi) n(E, \mathbf{r}, \tau_{\rm age})$. 

We employ the physically motivated ``Model A'' from our previous work~\cite{Zhang:2025tew} as the injection spectra. This model is characterized by a hard spectral shape with a steep cutoff. 
For comparison, we also evaluate a generic power-law injection spectrum with a spectral index of 2 and an exponential cutoff at 6~PeV~\cite{Aharonian:2026tzf}.

We parameterize transport using a parallel diffusion coefficient $D_\Vert = 6.1 \times 10^{28} (E / 4\text{ GeV})^{1/3} \text{ cm}^2 \text{ s}^{-1}$ along the regular magnetic field direction, with perpendicular diffusion defined by $D_{\perp} = \varepsilon D_{\parallel}$. Galactic CR grammage constraints favor $\varepsilon \sim 0.01\text{--}0.1$ for moderate-to-strong anisotropic regimes~\cite{Giacinti:2017dgt,AL-Zetoun:2020pfg}. 

We constrain the loading factor $\eta_{\rm cr}$ by fitting our model to the observed LHAASO CR proton bump~\cite{LHAASO:2025mlf}, which should be interpreted as the upper bound for the microquasar population. Transport parameters and GMF models are fixed at their nominal values to avoid normalization degeneracies with $\eta_{\rm cr}$, as these models are separately constrained by lower-energy GeV-TeV data. We adopt $\varepsilon=0.1$ as our baseline value and test $\varepsilon=0.01$ in the following section to assess the impact on local CR flux using a hypothetical nearby active source. Detailed propagation setups and additional physical impacts of these parameters are provided in Appendices~\ref{app:sim_setup} and \ref{app:diffusion}, respectively.

Throughout this work, we assume a cylindrical Galactic halo with radius $R_G = 20\text{ kpc}$ and half-height $H_G = 4\text{ kpc}$, with free escaping boundary for CRs~\cite{Maurin:2022gfm}.

\noindent\textit{GMF Effects}\label{sec:results}--
To contribute significantly ($\gtrsim 10\%$) to the observed PeV flux, a CR energy density of $\sim (0.4 - 4) \times 10^{-5}$ eV cm$^{-3}$ is required~\cite{LHAASO:2025byy}. Under the assumption of purely isotropic diffusion, using the method in Ref.~\cite{Blasi:2013rva}, we analytically calculate that a single source must be located within $d \lesssim 0.5$ kpc and have an age $\tau_{\rm age} \lesssim$ 0.2 Myr to dominate the local flux. As the distance increases to $d \sim 2$ kpc, the contribution to flux drops by approximately 90\%. 

\begin{figure}
    \centering\includegraphics[width=0.9\linewidth]{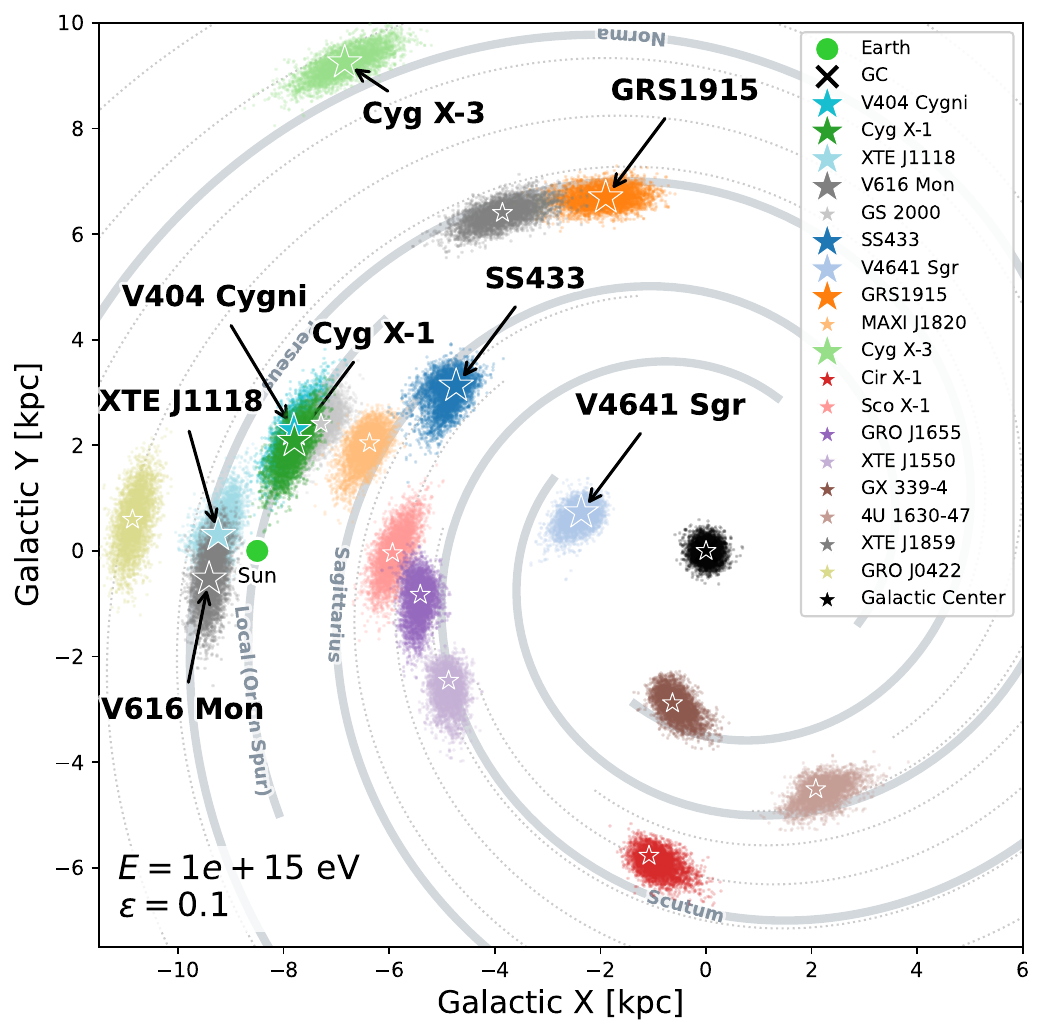}
        \vspace{-12pt}

    \caption{Illustration of the spatial distribution of 1 PeV cosmic rays (colored dots) from known Galactic microquasar locations (stars; Table~\ref{tab:mq}) under the GMF (gray spirals), assuming anisotropic diffusion of $\varepsilon = 0.1$.}
    \label{fig:cr-map-mq}
        \vspace{-10pt}
\end{figure}

However, the presence of GMF channels CRs along spiral arms via anisotropic diffusion~\cite{Jansson:2012pc} and makes magnetic connectivity a primary driver of the observed local flux. As illustrated in the spatial distribution of 1 PeV CRs in Fig.~\ref{fig:cr-map-mq}, this creates a filamentary morphology for the CR clouds. For sources at the positions of V616~Mon~\cite{Harrison:2006im} and XTE J1118+480~\cite{Gelino:2006sb}, the local magnetic field lines do not pass through the solar neighborhood. With inefficient cross-field transport ($\varepsilon \lesssim 0.1$), the particles remain confined to their original magnetic channels and are carried away from Earth. Consequently, their observed flux is suppressed despite their relative proximity.

Conversely, sources in the Cygnus region, such as Cyg~X-1~\cite{Miller-Jones:2021plh} and V404 Cygni~\cite{2009ApJ...706L.230M}, are magnetically connected to our location. This alignment enhances the observed flux by funneling their cosmic rays toward us, as illustrated in Fig.~\ref{fig:cr-map-mq}. We note locations as representative spatial markers to demonstrate flux enhancement, rather than a realization of their individual CR outputs.

To investigate the dominance of GMF over spatial proximity, we compare the predicted flux from the connected Cyg~X-1 ($d \approx 2.22$ kpc~\cite{Miller-Jones:2021plh}) with the non-connected V616~Mon ($d \approx 1.06$ kpc). For fair comparison, we inject CRs from a black hole of $10 M_\odot$ with $\tau_{\rm dur}=$0.2 Myr, assuming $\eta_{\rm cr} = 0.005$.

\begin{figure}
    \centering
    \includegraphics[width=\linewidth]{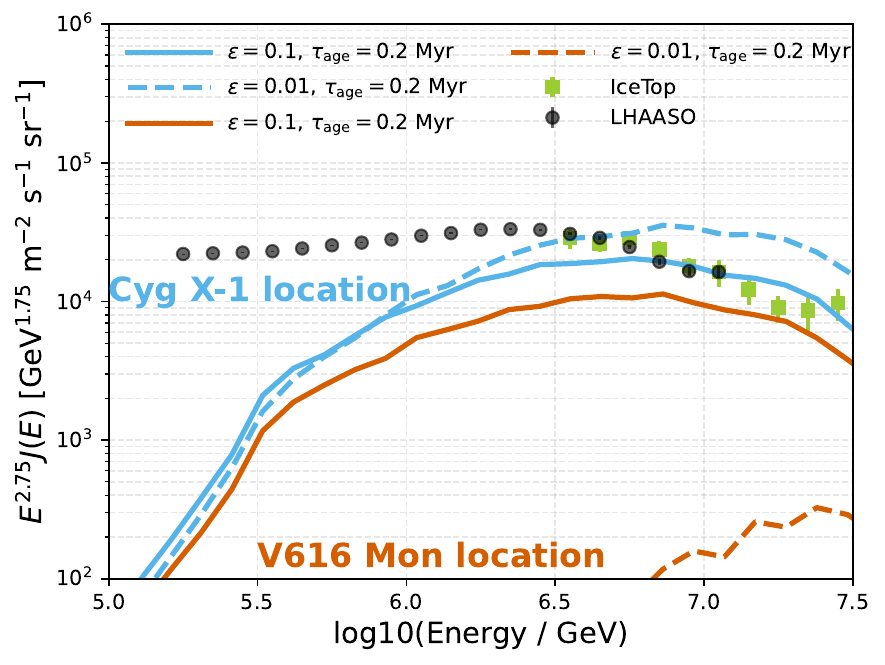}
    \includegraphics[width=\linewidth]{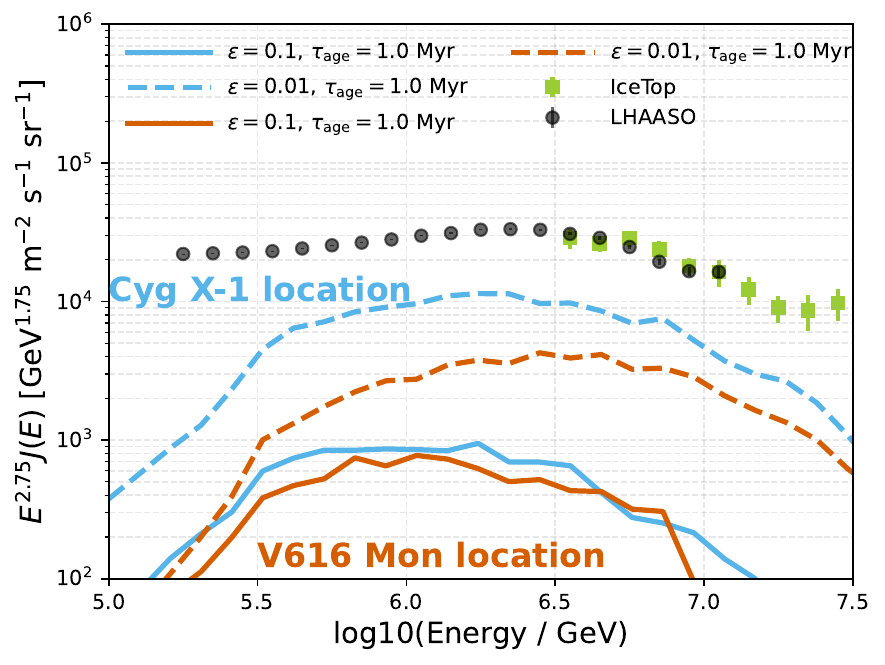}
    \vspace{-15pt}
    \caption{Predicted CR spectra from Cyg~X-1 and V616~Mon locations against data, comparing $\varepsilon = 0.1$ (solid), $0.01$ (dashed), $\tau_{\rm age} = 0.2$~Myr (top), and $1$~Myr (bottom), assuming Model A with $M_{\rm BH} = 10\,M_\odot$, $\tau_{\rm dur} = 0.2$~Myr, and $\eta_{\rm cr} = 0.005$.}
    \label{fig:cr-spectrum}
    \vspace{-10pt}
\end{figure}

The resulting fluxes in Fig.~\ref{fig:cr-spectrum} demonstrate that for an active source, the impact of the anisotropy diffusion parameter $\varepsilon$ (dashed versus solid lines) is more significant than the physical distance of the source (blue versus red curves). Because CRs from magnetically misaligned sources must undergo suppressed diffusive transport in the direction perpendicular to the field lines to reach Earth, the flux from V616~Mon is heavily attenuated. Consequently, the position of Cyg~X-1 represents a more viable source location for the observed PeV flux, despite its larger distance. As an active source on a connected GMF line, its flux remains robust even when $\varepsilon$ is reduced from 0.1 to 0.01.

The impact of anisotropic diffusion driven by the GMF is particularly pronounced for microquasar remnants. Once the central engines stopped injecting fresh CRs, the observable spatial distribution relies entirely on the magnetic field, making a source's position relative to the GMF as vital as its intrinsic energetic for shaping the ``knee". Comparing the active and remnant scenarios in Fig.~\ref{fig:cr-spectrum} illustrates a clear spectral evolution: active sources produce significantly harder spectra and a higher flux, indicating that active nearby microquasars can generate localized spectral features at high energies.

However, the absence of such features above 10 PeV in current observations suggests that CRs from currently active sources may remain confined within their local environments. Furthermore, the distribution around the Galactic Center in Fig.~\ref{fig:cr-map-mq} highlights the GMF confinement. In contrast, CRs from microquasar remnants have had sufficient time to escape their local environments and diffuse through the Galaxy, softening their spectra and filling the diffuse PeV CR ``sea'' discussed below.

\noindent\textit{Contributions of Microquasar Remnants}\label{sec:remnant_mq}--
Microquasars contribute non-negligibly to the observed local CR flux, which is governed by particle injection, transport, and escaping across the Galactic Halo boundary before significant energy losses occur. The $pp$ timescale, $\tau_{\text{loss}} \simeq 50 \left(n_{\text{gas}}/{1\,\text{cm}^{-3}}\right)^{-1} \text{Myr}$, corresponds to a vast source horizon $d_H = \sqrt{6Dt_{pp}} \sim 60$ kpc, ensuring that PeV protons escape the halo long before significant hadronic attenuation. 
However, CRs must also have sufficient time to diffuse to the Solar neighborhood after escaping from the remnants. While most distant microquasars are too short-lived to be individually resolved from Earth on Galactic timescales, they serve as primary injectors for the diffuse CR ``sea,'' with their remnant clouds continuing to expand and contribute long after the central engines shut off. Consequently, the local observed CR is shaped predominantly by these ``remnants'', representing an integrated history of Galactic microquasar activity.

To quantify these contributions, we uniformly sample Galactic microquasars over a 50 Myr history, assuming a constant birth rate $\mathcal{R}_{\rm MQ} \approx 1 \times 10^{-4} \text{ yr}^{-1}$~\cite{Podsiadlowski:2002ww}. This rate yields a current population of $\sim 50\text{--}60$ active microquasars, broadly consistent with the observed Galactic black hole X-ray binary candidates of $\sim 70\text{--}100$~\cite{2026arXiv260706360O}.
Each source is assigned an active lifetime between 0.1 and 1 Myr, a spatial distribution following the radial profile of supernova remnants~\cite{Green2015}, and black hole masses ($M_{\rm BH}$) sampled such that $\log(M_{\rm BH}/M_{\odot})$ ranges between 5 and 10. This yields an Eddington luminosity scale of $L_{\rm Edd} \simeq 1.26 \times 10^{39} (M_{\rm BH}/10~M_{\odot}) \text{ erg s}^{-1}$~\cite{1979rpa..book.....R}. For computational efficiency and to focus on sources contributing to the observed CR flux above 100 TeV, we simulate microquasars within 0.5-5 kpc and with ages up to 6.5 Myr. 

\begin{figure}
    \centering
    \includegraphics[width=\linewidth]{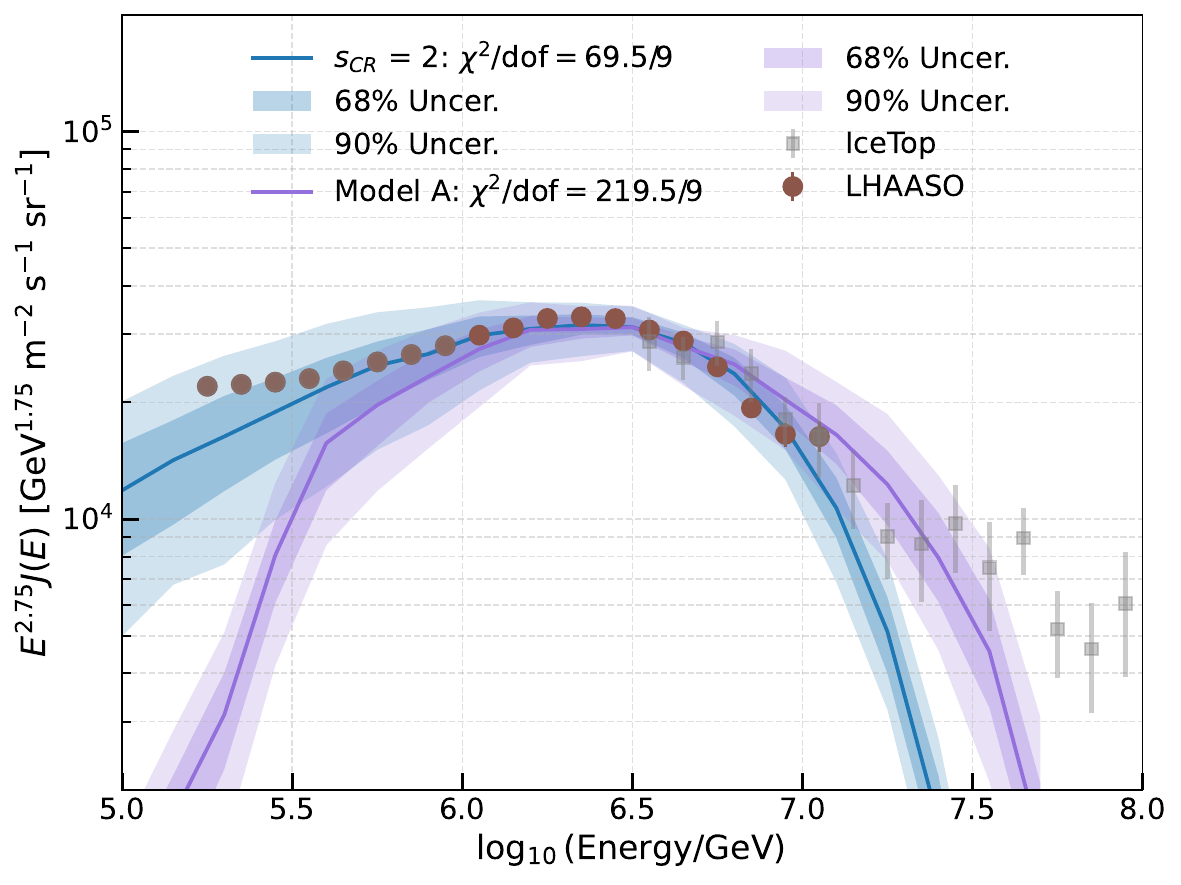}
    \vspace{-15pt}
    \caption{Diffuse CR energy spectra 68\% (darker shaded) and 95\% (lighter shaded) quantiles of the Galactic microquasar remnant realizations, assuming $\varepsilon = 0.1$, with Model A (violet) evaluated against power-law scenario (blue) compared against LHAASO (dot) and IceTop (square) data, with those (not) used in the fit colored in (gray) brown.}
    \label{fig:diffuse}
    \vspace{-15pt}
\end{figure}

The predicted energy spectra of simulated realizations are fitted to local CR proton data in the PeV bump region ([1, 10]~PeV) with the acceleration loading efficiency $\eta_{\rm cr}$ as a free parameter. The best-fit loading efficiencies are $\eta_{\rm cr} = 0.005_{-0.0025}^{+0.012}$ for Model A and $\eta_{\rm cr} = 0.03_{-0.015}^{+0.072}$ for the $E^{-2}$ power-law injection scenario ($s_{\rm cr} = 2$, $E_{\rm max} = 6~\text{PeV}$), both remaining well below the $\sim 10\% L_{\rm Edd}$ benchmark typical of jet kinetic power~\cite{2005Natur.436..819G}.
These results yield an integrated Galactic CR injection rate $Q_{\rm cr, tot} \simeq (2.0$--$9.8) \times 10^{38}\text{ erg s}^{-1}$ for Model A or $Q_{\rm cr, tot} \simeq (1.3$--$6.5) \times 10^{39}\text{ erg s}^{-1}$ ($E^{-2}$) for an $E^{-2}$ spectrum. These are comparable with the empirical Galactic volume-integrated rate of $\sim 1.5 \times 10^{38}\text{ erg s}^{-1}$~\cite{LHAASO:2025byy, Zhang:2025tew, Wang:2025yqy} when accounting for bolometric corrections.

As shown in Fig.~\ref{fig:diffuse}, the resulting $1\sigma$ percentile bands accommodate the observed PeV bump for both injection scenarios, with the latter extending smoothly down to $\sim 100~\text{TeV}$. These results indicate that the PeV bump in the CR proton spectrum arises naturally as an intrinsic feature of the microquasar population rather than being an occasional outlier, where stochastic spatial fluctuations govern the variance across realizations (see Fig.~\ref{fig:wideband} in Appendix~\ref{app:sim_setup}). Critically, an inherent degeneracy exists between local source distribution and CR acceleration efficiency: a modest loading efficiency $\eta_{\rm cr}$ can be offset by the proximity of nearby sources, whereas a higher efficiency permits more distant distributions. The best-fit value ($\eta_{\rm cr} \sim 0.03$) demonstrates that reproducing the observed PeV flux requires only modest jet acceleration efficiencies, establishing that local flux levels could be naturally accommodated by standard microquasar energetics.

\begin{figure}
       \centering
       \includegraphics[width=1.0\linewidth]{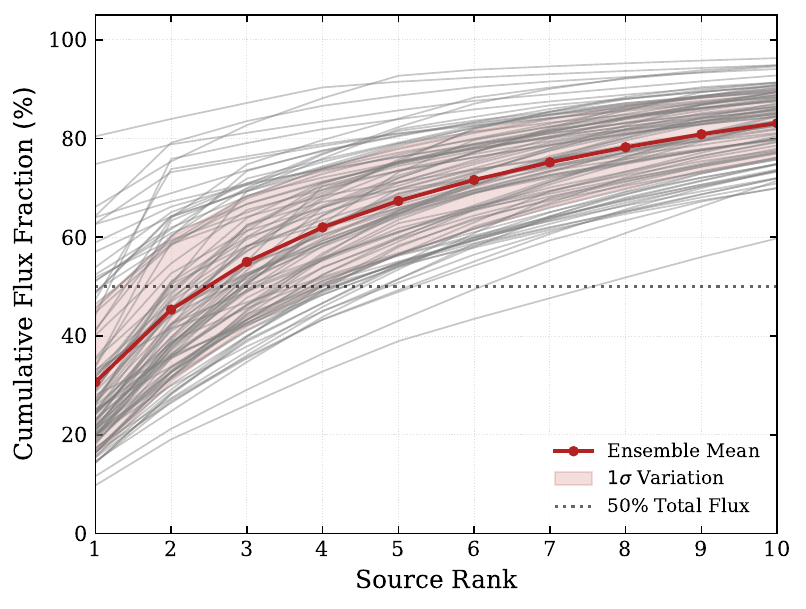}
        \vspace{-15pt}
\caption{Cumulative fraction of the local 1~PeV CR flux versus sources ranked by individual contribution (gray), median (solid red), and 68\% band (shade).}
       \label{fig:cumulative}
        \vspace{-15pt}
   \end{figure}

We calculate and rank the fractional flux contribution of individual sources in each realization, as shown in Fig.~\ref{fig:cumulative}. The 68\% interval of cumulative fractional flux across the ensemble indicates that typically no more than four sources are needed to account for 50\% of the observed flux at 1~PeV, while the top $\sim 10$ sources account for $\sim 80\%$.

Additionally, we note that the choice of Galactic halo size modestly affects the overall flux, as a larger halo traps and diffuses CRs from intermediate distances before they escape from the boundary (see Appendix~\ref{app:diffusion} for details).

\noindent\textit{Discussion and Conclusions}--
Evaluating the contribution of microquasars to local PeV CRs requires carefully accounting for the interplay between the injection spectrum, source age, and GMF transport. Our results demonstrate that both a hard and a standard $E^{-2}$ power-law injection spectrum can reproduce the observed PeV flux while maintaining acceptable energy budgets of $\eta_{\rm cr} \simeq 0.3$--$1.5\%$ and $\eta_{\rm cr} \simeq 2$--$10\%$, respectively. A harder spectrum therefore places a less stringent requirement on the energy budget. Since other CR sources such as supernova remnants (SNRs) and star-forming regions may also produce PeV protons~\cite{Aharonian:2026tzf}, the CR loading efficiency for microquasars should not exceed our constraints, keeping the necessary energy input well within plausible jet energetics.

Rather than reflecting a continuous global average, the local PeV CR flux is driven predominantly by a small number of recent, nearby microquasar remnants drawn from the broader Galactic CR ``sea''. To maintain agreement with high-energy data, freshly injected CRs from active sources may still be in transit toward Earth or locally confined near the accelerator~\cite{Ohira:2024qtr}, reducing their present contribution to the local CR spectrum. Evaluating potential confinement within extended local environments, such as microquasar lobes~\cite{2011A&A...528A..89B}, remains critical for characterizing both active systems and recent remnants. In microquasar remnants, such confinement effects diminish as particles escape into interstellar space. However, active sources could be far more prominent in multi-messenger signals, as prompt hadronic interactions produce neutrinos and gamma rays that escape unaffected by magnetic fields. While our main focus is on protons, heavier CR species accelerated at equal rigidity yield similar energy spectra, preserving overall composition trends around the ``knee''~\cite{Zhang:2025tew}.

Alternative propagation scenarios, such as space-dependent diffusion motivated by recent boron-to-carbon measurements~\cite{2022SciBu..67.2162D}, updated coherent GMF models~\cite{Unger:2023lob}, and ballistic transport beyond a few tens of PeV, may further refine local flux predictions and should be considered in future work.

In summary, our framework demonstrates that the local CR environment at PeV energies can be feasibly interpreted through the accumulated output of microquasar remnants. The PeV sky is therefore a signature of the integrated history of the Galaxy, where the contributions of ancient jets persist long after their central engines have ceased. Other PeVatron candidates, such as specific supernova remnants~\cite{Ptuskin2012}, pulsar wind nebulae~\cite{LHAASO:2021cbz}, and the supermassive black hole at the Galactic Center~\cite{HESS:2016pst}, remain potential contributors; the framework proposed in this work could also be used to test the associations of these PeVatron candidates to the observed CR ``knee''. In a recent follow-up study, we extend this framework and numerically calculate the associated secondary neutrino and gamma-ray emission, providing testable multi-messenger predictions anchored in CR data~\cite{Yu:2026vco}.

\clearpage

\appendix
\renewcommand{\thetable}{S\arabic{table}}
\renewcommand{\thefigure}{S\arabic{figure}}

\section{\textbf{Microquasar catalogs}}\label{mq_cat}

Table~\ref{tab:mq} summarizes the spatial parameters of the observed microquasars used to generate the map in Fig.~\ref{fig:cr-map-mq}.

\begin{table}
    \caption{Galactic coordinates and distances for some microquasars, ordered by proximity to the Sun. The Galactic Center is provided for reference.}
    \centering
    \label{tab:mq}
    \begin{threeparttable}
    \begin{tabular}{l c c c}
        \hline
        \hline
        \textbf{Source Name} & \textbf{$l$} & \textbf{$b$} & \textbf{$d$} \\
         & ($^{\circ}$) & ($^{\circ}$) & (kpc) \\
        \hline
        \hline
        V616~Mon~\cite{Harrison:2006im}          & $210.04$ & $-6.54$  & $1.06$ \\
        XTE J1118+480~\cite{Gelino:2006sb}     & $157.69$ & $+62.31$ & $1.72$ \\
        Cygnus X-1~\cite{Miller-Jones:2021plh}        & $71.33$  & $+3.07$  & $2.22$ \\
        V404 Cygni~\cite{2009ApJ...706L.230M}        & $73.12$  & $-2.07$  & $2.39$ \\
        GRO J0422+32~\cite{Gelino:2003pr}      & $165.88$ & $-11.91$ & $2.49$ \\
        GS 2000+251~\cite{1996ApJ...470L..57C}       & $63.37$  & $-3.01$  & $2.70$ \\
        Scorpius X-1~\cite{1999ApJ...512L.121B}  & $359.09$ & $+23.78$ & $2.80$ \\
        MAXI J1820+070~\cite{Atri:2019kws}  & $43.83$  & $+5.29$  & $2.96$ \\
        GRO J1655$-$40~\cite{Hjellming:1995tv}  & $344.98$ & $+2.46$  & $3.20$ \\
        4U 1755$-$33~\cite{Angelini:2003sj}      & $357.22$ & $-4.87$  & $4.00$ \\
        CI Cam~\cite{2023AstBu..78....1B}            & $148.47$ & $+4.26$  & $4.10$ \\
        XTE J1550$-$564~\cite{Migliori:2017jqq} & $325.88$ & $-1.84$  & $4.38$ \\
        SS 433~\cite{2018ApJ...863..103S}            & $39.70$  & $-2.20$  & $4.90$ \\
        V4641 Sgr~\cite{MacDonald:2014gpa}         & $6.77$   & $-4.79$  & $6.20$ \\
        XTE J1859+226~\cite{2005ApJ...623.1026H}     & $54.05$  & $+8.61$  & $8.00$ \\
        GX 339$-$4~\cite{Parker:2016ltr}        & $339.86$ & $-4.32$  & $8.40$ \\
        \textit{Galactic Center} & $0.00$   & $0.00$   & $8.50$ \\
        GRS 1915+105~\cite{Inokuchi:2025mch, reid2023distances}      & $45.40$  & $-0.20$  & $9.40$ \\
        Cygnus X-3~\cite{Reid:2023ksq}        & $79.85$  & $+0.68$  & $9.40$ \\
        Circinus X-1~\cite{Heinz:2015jla}  & $322.12$ & $-0.03$    & $9.40$ \\
        4U 1630$-$47~\cite{Kalemci:2018tkp}      & $336.91$ & $+0.25$  & $11.50$ \\
        \hline
        \hline
    \end{tabular}
    \end{threeparttable}
\end{table}

\begin{figure}[bth]
    \centering
    \includegraphics[width=\linewidth]{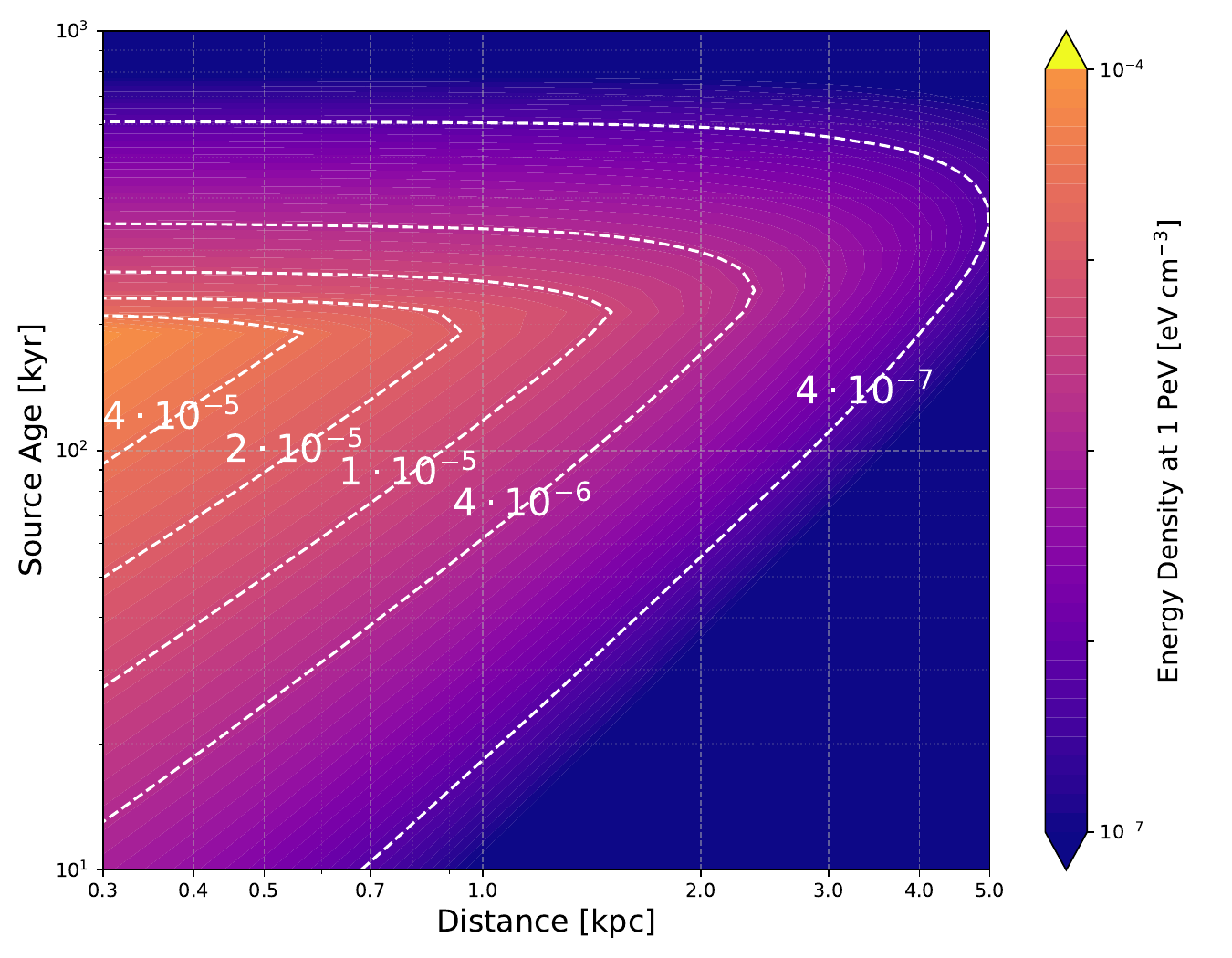}
        \vspace{-12pt}

    \caption{Cosmic ray energy density at 1 PeV shown as a function of distance and source age. The model assumes a jet duration $\tau_{\rm dur} = 0.2$ Myr from a $10 M_\odot$ black hole and isotropic diffusion without GMF effects. 
    The injection spectrum $Q(E)$ is normalized such that the total cosmic ray luminosity is $L_{\rm cr} = \eta_{\rm cr} L_{\rm Edd}$, where we adopt $\eta_{\rm cr} = 0.1$. The spectral index is assumed to be $s_{\rm cr} = 2$.
    Contours identify density levels between $4 \times 10^{-7}$ and $4 \times 10^{-5}$ eV cm$^{-3}$.}
    \label{fig:contour-PeV}
    \vspace{-10pt}

\end{figure}

The parameter space illustrated in Fig.~\ref{fig:contour-PeV} quantifies the conditions under which an individual source might dominate the local PeV flux. The central contours indicate that dominance is restricted to nearby, recently active sources ($d \lesssim 0.5$ kpc, $\tau_{\rm age} \lesssim 0.2$ Myr). At greater distances ($d \sim 2$ kpc), the contribution falls by nearly an order of magnitude. These results contextualize the values for known objects like Cygnus X-1, which has a distance $d = 2.22$ kpc~\cite{Miller-Jones:2021plh} and an estimated jet duration of 0.04 to 0.3 Myr~\cite{2005Natur.436..819G}. While such sources contribute minimally under the assumption of isotropic diffusion without GMF, their role is significantly modulated when accounting for the magnetic field.

\section{CR propagation and numerical Simulation Setup}\label{app:sim_setup}
The numerical simulation of CR injection, propagation, and transportation under the Galactic magnetic field is performed using {\sc CRPropa 3.2}. 
To track the diffusive evolution of the particle density distribution of each primary proton within the solenoidal JF-2012 field configuration (\texttt{JF12FieldSolenoidal}), the \texttt{DiffusionSDE} module is utilized to stochastically solve the equivalent microscopic Stochastic Differential Equation (SDE).
The underlying algorithmic implementation and verification of this numerical transport framework are established in~\cite{CRPropa:2022ovg}. To accurately resolve the structural gradients of these field components when transitioning throughout the full GMF model, the spatial integration step size is dynamically constrained between a minimum threshold of $10\text{ pc}$ and a maximum limit of $100\text{ pc}$.

For each source, $10^5$ monochromatic primary proton pseudo-particles are injected from the simulated microquasar positions per bin across \textcolor{blue}{31} logarithmic bins in the energy range of $10^{14}\text{--}10^{17}\text{ eV}$. 
The free escape boundary conditions centered at the Galactic origin are handled by the \texttt{CylindricalBoundary} module, configured with a radius of $20\text{ kpc}$ and a half-height of $2\text{ kpc}$ along the Galactic plane. Particles crossing this surface envelope are immediately deactivated and removed from the active tracking stack. For effective simulation, we utilize the \texttt{MaximumTrajectoryLength} module to enforce a maximum trajectory threshold defined by the maximum lookback time ($l_{\text{max}} = c \cdot \tau_{\text{age}}$) and terminate individual paths once they exceed the threshold, where $\tau_{\text{age}}$ is the lookback time of the source.
Within the global \texttt{Observer}, the \texttt{ObserverTimeEvolution} module is used to track the spatial coordinates of simulated particles over $3000$ discrete snapshot intervals $\Delta l = c \Delta \tau = c\cdot \tau_{\text{dur}}/3000$, with an initial minimum trajectory length threshold ($l_{\text{min}} = c \cdot [\tau_{\text{age}} - \tau_{\text{dur}}]$), where $t_{\text{dur}}$ denotes the physical duration of the source. 
The observer detection flags are explicitly configured to prevent particle deactivation upon recording (\texttt{setDeactivateOnDetection(False)}), preserving the continuous evolution of the population across all active snapshot epochs.

To evaluate the time-integrated density $n(E, \mathbf{r}, \tau_{\rm age})$, the propagator $\mathcal{P}$ is estimated from the SDE ensemble at each discrete snapshot epoch. This is achieved by filtering for the subset of particles $N_{\text{inside}}$ that intersect a local spherical observer volume ($V_{\oplus} = \frac{4}{3}\pi R_{\text{obs}}^3$) centered at the Solar position $\mathbf{r}_{\oplus} = (-8.5\,\text{kpc}, 0, 0)$ with a radius of $R_{\text{obs}} = 200\text{ pc}$:\begin{equation}\mathcal{P}(\mathbf{r}, E, \tau) \approx \frac{1}{N_{\text{particles}} V_{\oplus}} \sum_{i=1}^{N_{\text{particles}}} \delta_{V}(\mathbf{r}_i(\tau)) = \frac{N_{\text{inside}}(\tau)}{N_{\text{particles}} V_{\oplus}},\end{equation} where $\delta_{V}$ acts as an indicator function that equals unity if the $i$-th simulated particle is inside $V_{\oplus}$ at lookback time $\tau$. The observed flux is estimated as:
\begin{equation}
I(E, t) = \frac{N_{\text{inside}} \cdot \Delta \tau}{V_{\oplus} \cdot N_{\text{particles}}} Q(E) c.
\end{equation}

The free escape boundary conditions centered at the Galactic origin are handled by the \texttt{CylindricalBoundary} module, configured with a radius of $20\text{ kpc}$ and a half-height of $4\text{ kpc}$ along the Galactic plane. Particles crossing this surface envelope are immediately deactivated and removed from the active tracking stack. For effective simulation, we utilize the \texttt{MaximumTrajectoryLength} module to enforce a maximum trajectory threshold defined by the maximum lookback time ($l_{\text{max}} = c \cdot t_{\text{obs}}$) and terminate individual paths once they exceed the threshold.
To account for spatial and temporal variance in the local microquasar population, we simulate an ensemble of 200 MC realizations using independent random seeds. The parameters governing catalog generation and particle injection are summarized in Table~\ref{tab:mq_params}.

\begin{table}[b]
\caption{\label{tab:mq_params}Parameters used in the microquasar catalog generation and CRPropa simulation setup.}
\begin{ruledtabular}
\begin{tabular}{ll}
Parameter (Symbol) & Value / Distribution \\
\hline
\multicolumn{2}{c}{\textit{Constants}} \\
Birth Rate & $1 \times 10^{-4}\text{ yr}^{-1}$ \\
CR Loading Efficiency Baseline ($\eta_{\text{cr}}$) & $0.1$ \\
Simulation Cutoff  & $6\text{ Myr}$ \& non-active \\
\hline
\multicolumn{2}{c}{\textit{Sample Ensembles}} \\
Lookback Time ($t_{\text{lookback}}$) & $\mathcal{U}(0, 50)\text{ Myr}$ \\
Jet Duration ($t_{\text{duration}}$) & $\mathcal{U}(0.1, 1.0)\text{ Myr}$ \\
Black Hole Mass ($M_{\text{BH}}$) & $\mathcal{U}(5, 15)\ M_\odot$ \\
Galactocentric Radius Distribution($r_{\text{gc}}$) & Green15\\
Vertical Height ($z_{\text{gc}}$) & $\pm\text{Exp}(h_z=0.15~\text{kpc})$ \\
\hline
\multicolumn{2}{c}{\textit{CRPropa Simulation}} \\
Injected Primary & Protons\\
Energy Range & [$100\text{ TeV}$, $100\text{ PeV}$] \\
Energy Bins & $21$ (log-spaced) \\
Particles per Bin & $25,000$ \\
Observer radius ($R_{\rm obs}$) & 200 pc \\
\end{tabular}
\end{ruledtabular}
\end{table}

\begin{figure}[htbp]\centering\includegraphics[width=\linewidth]{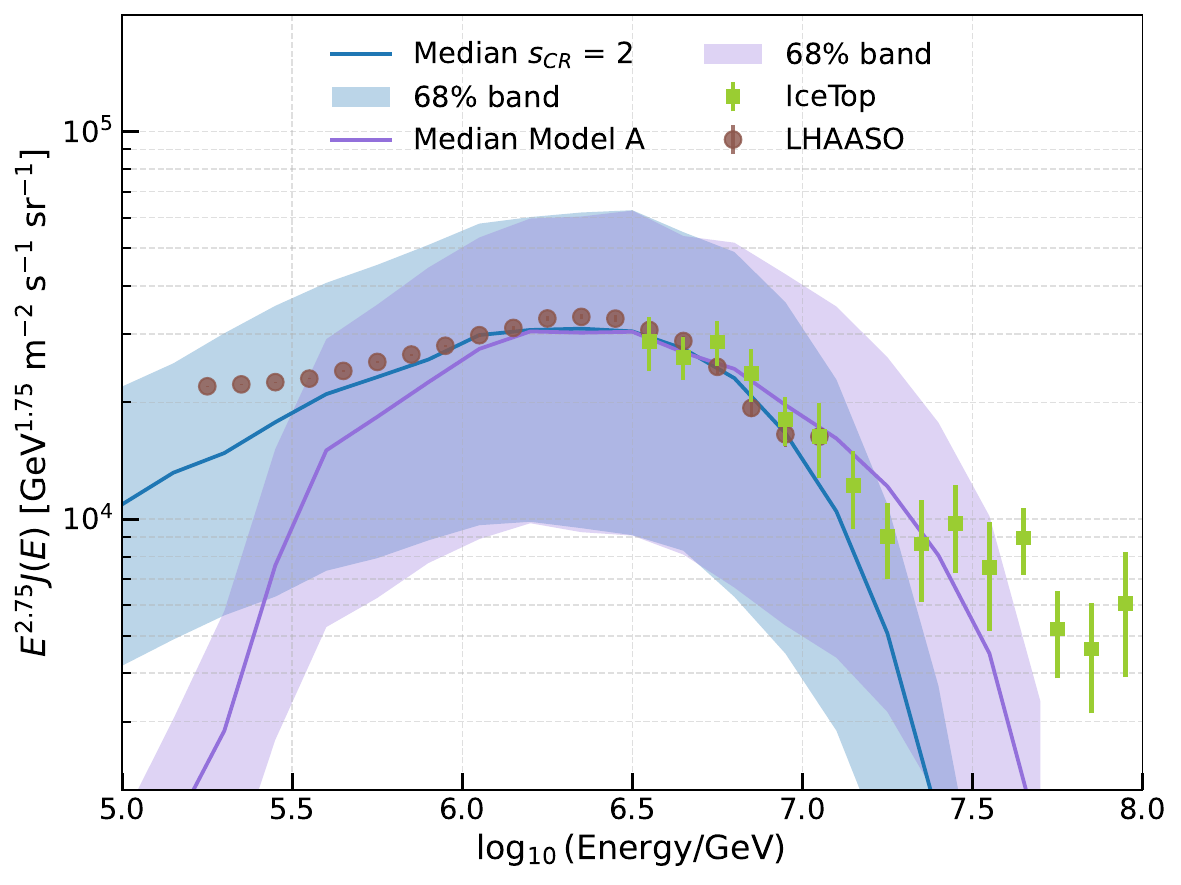}
    \vspace{-12pt}

\caption{Energy spectrum of CR protons resulting from simulated microquasar populations with best-fit values of $\eta_{cr}$ normalizing the median spectrum to local CR observations in the $[1, 10]\text{~PeV}$ interval. The shaded band denotes the 68\% quantiles across MC realizations, illustrating the intrinsic ensemble variance induced by the local source distribution.}\label{fig:wideband}
    \vspace{-10pt}
\end{figure}

The stochastic uncertainty driven by variations in the local source distribution is shown in Fig.~\ref{fig:wideband}, where the median curve across all realizations is scaled to match the local CR data in $[1, 10]\text{~PeV}$.

\section{Anisotropic diffusion and observational constraints }\label{app:diffusion}
The Galactic magnetic field consists of a large-scale regular component aligned with the spiral arms and a turbulent component spanning spatial scales from $l_{\rm max} \sim 10\text{ to }100~\text{pc}$ down to dissipation scales $l_{\rm min} \sim 1~\text{AU}$. 
In the presence of a regular guiding field, CR transport is naturally anisotropic ($D_\parallel > D_\perp$), causing propagation to be preferentially elongated along ordered field lines and directly shaping the spatial morphology of diffuse emission (Fig.~\ref{fig:cr-map-mq}). 
Our simulations incorporate both regular and turbulent fields using the \texttt{DiffusionSDE} module in \textsc{CRPropa~3.2}, adopting the coherent field from the JF12 model. The module defines the local guiding vector at each step, and tracks transport along it, encoding field-line random walk and turbulent fluctuations into the anisotropic diffusion tensor via $D_\perp = \varepsilon D_\parallel$. This formulation efficiently captures turbulent transport while maintaining computational efficiency over Galactic-scale transport. 

Our baseline choice of $\varepsilon = 0.1$ adopts the parameters established by detailed test-particle simulations constrained by secondary-to-primary ratios (e.g., boron-to-carbon ratio; B/C)~\cite{Giacinti:2017dgt, AL-Zetoun:2020pfg}. These studies indicate a dominant regular magnetic field component ($B_{\rm turb}/B_{\rm ord} \lesssim 1$), yielding an inferred anisotropy ratio of $\varepsilon = D_\perp / D_\parallel \sim 0.01\text{--}0.1$ for PeV protons.

Regarding $D_0$, we calculate its normalization based on observational constraints from B/C ratio, which constrains $D/H_{\rm G} \sim 0.03\text{--}0.1\,\text{kpc}\,\text{Myr}^{-1}$ at $10\,\text{GV}$~\cite{Genolini:2019ewc, Unger:2023lob}, while radioactive secondaries (${}^{10}\text{Be}/{}^9\text{Be}$) independently determine the halo size as $H_{\rm G} = 4.7 \pm 0.6 \pm 2.0\,\text{kpc}$~\cite{Johannesson:2016rlh, Maurin:2022gfm}. Taking $H_{\rm G} \sim 4\text{--}5\,\text{kpc}$ with a diffusion index $\delta = 1/3$, this corresponds to a baseline range of $D_0 \sim (3\text{--}9) \times 10^{28}\,\text{cm}^2\,\text{s}^{-1}$ at $E_0 = 4\,\text{GeV}$.

\begin{figure}
    \centering
    \includegraphics[width=\linewidth]{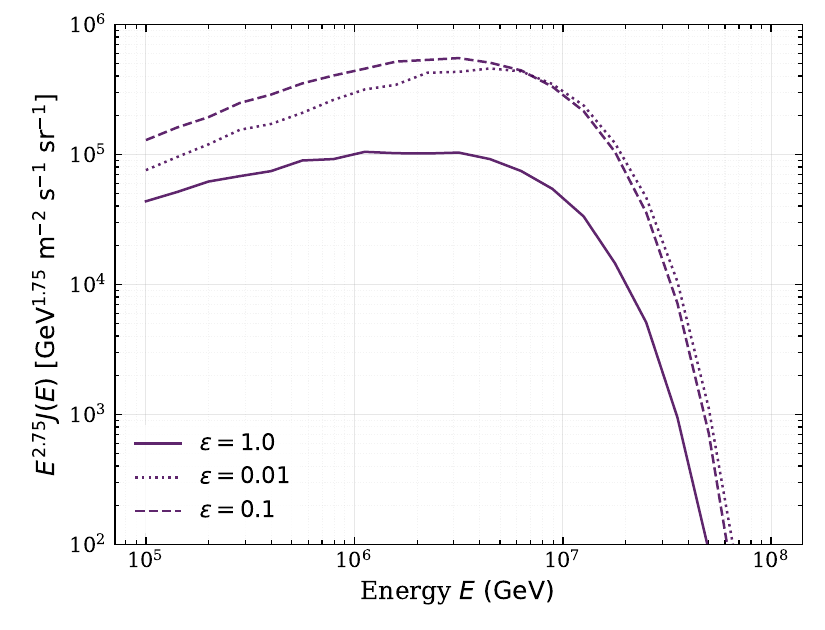}
        \vspace{-12pt}
    \caption{Predicted CR energy spectra for $\varepsilon = 1$, $0.1$, and $0.01$, respectively.}
    \label{fig:epsilon}
        \vspace{-10pt}
\end{figure}

To evaluate the impact of the anisotropy parameter, we calculate the diffuse CR flux across three benchmark levels ($\varepsilon = 1, 0.1, 0.01$), shown in Fig.~\ref{fig:epsilon}. The spectral shape around the PeV bump is preserved across anisotropic scenarios, with $\varepsilon$ primarily modifying the overall flux normalization and showing only a minor effect at higher energies. 

\section{Impact of Source Distribution Model}

\begin{figure}[hbt]
    \centering
    \includegraphics[width=\linewidth]{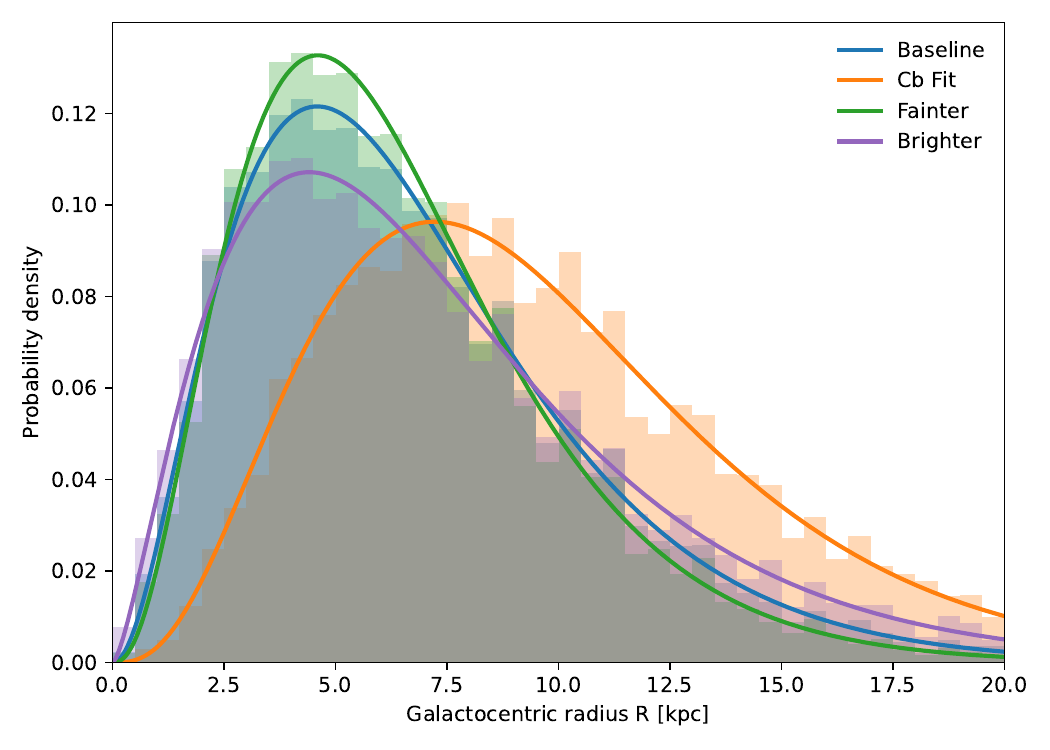}
    \includegraphics[width=\linewidth]{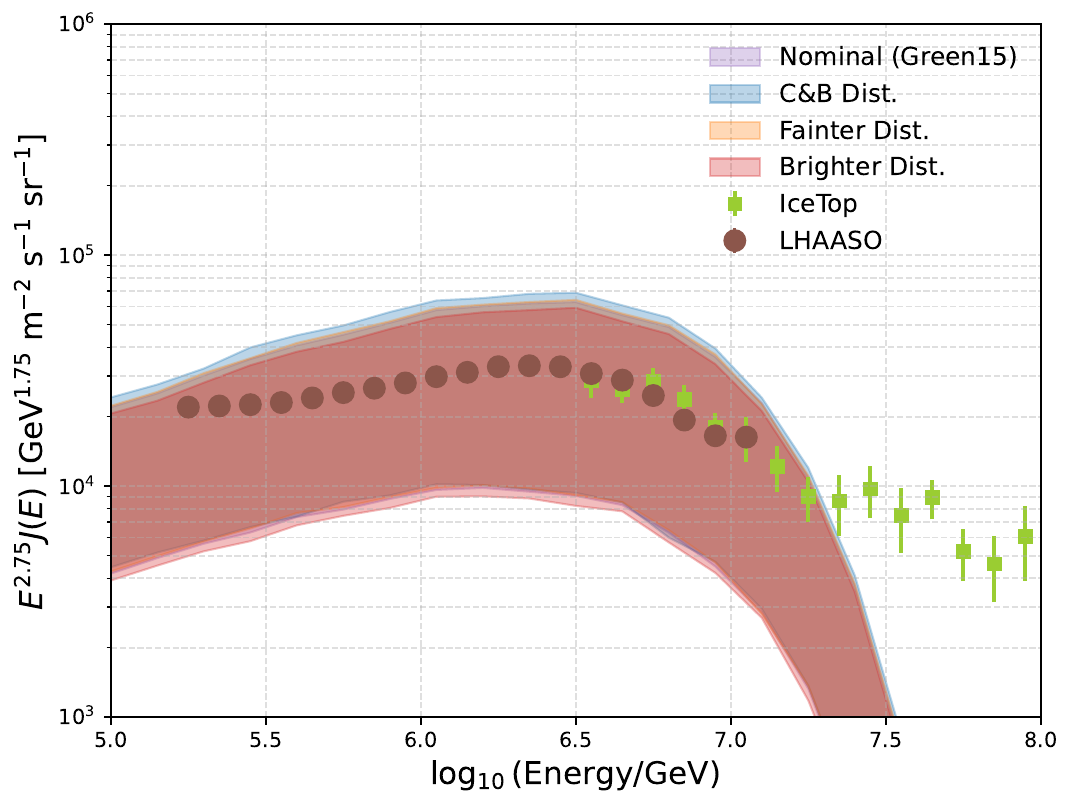}
    
    \caption{Top: Radial source distribution profiles (lines) and corresponding sampled microquasar ensembles. Bottom: Comparison of the predicted local CR spectra following four radial distributions.}
    \label{fig:radial_distribution}
\end{figure}

To evaluate the sensitivity of our predictions to the source distribution models, we evaluate three additional radial distribution models commonly adopted in the literature~\cite{Green2015,Case:1998qg} (Fig.~\ref{fig:radial_distribution}, top). As shown in the lower panel of Fig.~\ref{fig:radial_distribution}, variations in the spatial distribution yield negligible impacts on the predicted CR flux at PeV energies. 

\clearpage
\bibliography{main}

@PREAMBLE{
 "\providecommand{\noopsort}[1]{}" 
 # "\providecommand{\singleletter}[1]{#1}%" 
}

@BOOK{1979rpa..book.....R,
       author = {{Rybicki}, George B. and {Lightman}, Alan P.},
        title = "{Radiative processes in astrophysics}",
         year = 1979,
       adsurl = {https://ui.adsabs.harvard.edu/abs/1979rpa..book.....R}
}

@article{CRPropa:2022ovg,
    author = "Alves Batista, Rafael and others",
    collaboration = "CRPropa",
    title = "{CRPropa 3.2 {\textemdash} an advanced framework for high-energy particle propagation in extragalactic and galactic spaces}",
    eprint = "2208.00107",
    archivePrefix = "arXiv",
    primaryClass = "astro-ph.HE",
    doi = "10.1088/1475-7516/2022/09/035",
    journal = "JCAP",
    volume = "09",
    pages = "035",
    year = "2022"
}

@article{Jansson:2012rt,
    author = "Jansson, Ronnie and Farrar, Glennys R.",
    title = "{The Galactic Magnetic Field}",
    eprint = "1210.7820",
    archivePrefix = "arXiv",
    primaryClass = "astro-ph.GA",
    doi = "10.1088/2041-8205/761/1/L11",
    journal = "Astrophys. J. Lett.",
    volume = "761",
    pages = "L11",
    year = "2012"
}

@article{Blasi:2013rva,
    author = "Blasi, Pasquale",
    title = "{The Origin of Galactic Cosmic Rays}",
    eprint = "1311.7346",
    archivePrefix = "arXiv",
    primaryClass = "astro-ph.HE",
    doi = "10.1007/s00159-013-0070-7",
    journal = "Astron. Astrophys. Rev.",
    volume = "21",
    pages = "70",
    year = "2013"
}

@unpublished{LHAASO:2025byy,
    author = "Cao, Zhen and others",
    collaboration = "LHAASO",
    title = "{First Identification and Precise Spectral Measurement of the Proton Component in the Cosmic-Ray `Knee'}",
    eprint = "2505.14447",
    archivePrefix = "arXiv",
    primaryClass = "astro-ph.HE",
    month = "5",
    year = "2025"
}

@article{LHAASO:2024psv,
    author = "Cao, Zhen and others",
    collaboration = "LHAASO",
    title = "{Ultrahigh-energy gamma-ray emission associated with black hole{\textendash}jet systems}",
    eprint = "2410.08988",
    archivePrefix = "arXiv",
    primaryClass = "astro-ph.HE",
    doi = "10.1093/nsr/nwaf496",
    journal = "Natl. Sci. Rev.",
    volume = "12",
    number = "12",
    pages = "nwaf496",
    year = "2025"
}

@article{LHAASO:2024knt,
    author = "Cao, Zhen and others",
    collaboration = "LHAASO",
    title = "{Measurements of All-Particle Energy Spectrum and Mean Logarithmic Mass of Cosmic Rays from 0.3 to 30~PeV with LHAASO-KM2A}",
    eprint = "2403.10010",
    archivePrefix = "arXiv",
    primaryClass = "astro-ph.HE",
    doi = "10.1103/PhysRevLett.132.131002",
    journal = "Phys. Rev. Lett.",
    volume = "132",
    number = "13",
    pages = "131002",
    year = "2024"
}

@article{Zhang:2025tew,
    author = "Zhang, B. Theodore and Kimura, Shigeo S. and Murase, Kohta",
    title = "{Microquasar jet-cocoon systems as PeVatrons}",
    eprint = "2506.20193",
    archivePrefix = "arXiv",
    primaryClass = "astro-ph.HE",
    doi = "10.1103/p6r1-qg5q",
    journal = "Phys. Rev. D",
    volume = "112",
    number = "12",
    pages = "123015",
    year = "2025"
}

@article{Wang:2025yqy,
    author = "Wang, Jieshuang and Reville, Brian and Aharonian, Felix A.",
    title = "{Galactic Superaccreting X-Ray Binaries as Super-PeVatron Accelerators}",
    eprint = "2507.21048",
    archivePrefix = "arXiv",
    primaryClass = "astro-ph.HE",
    doi = "10.3847/2041-8213/adf3a4",
    journal = "Astrophys. J. Lett.",
    volume = "989",
    number = "2",
    pages = "L25",
    year = "2025"
}

@article{Harrison:2006im,
    author = "Harrison, Thomas E. and Howell, Steve B. and Szkody, Paula and Cordova, France A.",
    title = "{The Nature of the Secondary Star in the Black Hole X-Ray Transient V616 Mon (=A0620-00)}",
    eprint = "astro-ph/0609535",
    archivePrefix = "arXiv",
    doi = "10.1086/509572",
    journal = "Astron. J.",
    volume = "133",
    pages = "162--168",
    year = "2007"
}

@article{Miller-Jones:2021plh,
    author = "Miller-Jones, James C. A. and others",
    title = "{Cygnus X-1 contains a 21{\textendash}solar mass black hole{\textemdash}Implications for massive star winds}",
    eprint = "2102.09091",
    archivePrefix = "arXiv",
    primaryClass = "astro-ph.HE",
    doi = "10.1126/science.abb3363",
    journal = "Science",
    volume = "371",
    number = "6533",
    pages = "1046--1049",
    year = "2021"
}

@ARTICLE{2009ApJ...706L.230M,
       author = {{Miller-Jones}, J.~C.~A. and {Jonker}, P.~G. and {Dhawan}, V. and {Brisken}, W. and {Rupen}, M.~P. and {Nelemans}, G. and {Gallo}, E.},
        title = "{The First Accurate Parallax Distance to a Black Hole}",
      journal = {\apjl},
         year = 2009,
        month = dec,
       volume = {706},
       number = {2},
        pages = {L230-L234},
          doi = {10.1088/0004-637X/706/2/L230},
archivePrefix = {arXiv},
       eprint = {0910.5253},
 primaryClass = {astro-ph.HE},
       adsurl = {https://ui.adsabs.harvard.edu/abs/2009ApJ...706L.230M}
}

@ARTICLE{1996ApJ...470L..57C,
       author = {{Callanan}, Paul J. and {Garcia}, Michael R. and {Filippenko}, Alexei V. and {McLean}, Ian and {Teplitz}, Harry},
        title = "{On the Mass of the Black Hole in GS 2000+25}",
      journal = {\apjl},
         year = 1996,
        month = oct,
       volume = {470},
        pages = {L57},
          doi = {10.1086/310287},
       adsurl = {https://ui.adsabs.harvard.edu/abs/1996ApJ...470L..57C}
}

@article{Atri:2019kws,
    author = "Atri, P. and others",
    title = "{A radio parallax to the black hole X-ray binary MAXI J1820+070}",
    eprint = "1912.04525",
    archivePrefix = "arXiv",
    primaryClass = "astro-ph.HE",
    doi = "10.1093/mnrasl/slaa010",
    journal = "Mon. Not. Roy. Astron. Soc.",
    volume = "493",
    number = "1",
    pages = "L81--L86",
    year = "2020"
}

@article{Reid:2023ksq,
    author = "Reid, M. J. and Miller-Jones, J. C. A.",
    title = "{On the Distances to the X-Ray Binaries Cygnus X-3 and GRS 1915+105}",
    eprint = "2309.15027",
    archivePrefix = "arXiv",
    primaryClass = "astro-ph.HE",
    doi = "10.3847/1538-4357/acfe0c",
    journal = "Astrophys. J.",
    volume = "959",
    number = "2",
    pages = "85",
    year = "2023"
}

@article{Gelino:2006sb,
    author = "Gelino, Dawn M. and Balman, Solen and Kizilouglu, Umit and Yilmaz, Arda and Kalemci, Emrah and Tomsick, John A.",
    title = "{The inclination angle of and mass of the black hole in xte j1118+480}",
    eprint = "astro-ph/0601409",
    archivePrefix = "arXiv",
    doi = "10.1086/500924",
    journal = "Astrophys. J.",
    volume = "642",
    pages = "438--442",
    year = "2006"
}

@article{Gelino:2003pr,
    author = "Gelino, Dawn M. and Harrison, Thomas E.",
    title = "{Gro j0422+32: the lowest mass black hole?}",
    eprint = "astro-ph/0308490",
    archivePrefix = "arXiv",
    doi = "10.1086/379311",
    journal = "Astrophys. J.",
    volume = "599",
    pages = "1254--1259",
    year = "2003"
}

@ARTICLE{1999ApJ...512L.121B,
       author = {{Bradshaw}, C.~F. and {Fomalont}, E.~B. and {Geldzahler}, B.~J.},
        title = "{High-Resolution Parallax Measurements of Scorpius X-1}",
      journal = {\apjl},
         year = 1999,
        month = feb,
       volume = {512},
       number = {2},
        pages = {L121-L124},
          doi = {10.1086/311889},
       adsurl = {https://ui.adsabs.harvard.edu/abs/1999ApJ...512L.121B}
}

@article{Heinz:2015jla,
    author = "Heinz, Sebastian and Burton, Michael G. and Braiding, Catherine and Brandt, William N. and Jonker, Peter G. and Sell, Paul and Fender, Robert P. and Nowak, Michael A. and Schulz, Norbert S.",
    title = "{Lord of the Rings: A Kinematic Distance to Circinus X-1 from a Giant X-Ray Light Echo}",
    eprint = "1506.06142",
    archivePrefix = "arXiv",
    primaryClass = "astro-ph.HE",
    doi = "10.1088/0004-637X/806/2/265",
    journal = "Astrophys. J.",
    volume = "806",
    number = "2",
    pages = "265",
    year = "2015"
}

@article{Hjellming:1995tv,
    author = "Hjellming, R. M. and Rupen, M. P.",
    title = "{Episodic Ejection of Relativistic Jets by the X-Ray Transient GRO:J1655-40}",
    doi = "10.1038/375464a0",
    journal = "Nature",
    volume = "375",
    pages = "464--468",
    year = "1995"
}

@article{Migliori:2017jqq,
    author = {Migliori, Giulia and Corbel, St{\'e}phane and Tomsick, John A. and Kaaret, Philip and Fender, Rob P. and Tzioumis, Tasso and Coriat, Micka{\"e}l and Orosz, Jerome A.},
    title = "{Evolving morphology of the large-scale relativistic jets from XTE J1550{\ensuremath{-}}564}",
    eprint = "1707.06876",
    archivePrefix = "arXiv",
    primaryClass = "astro-ph.HE",
    doi = "10.1093/mnras/stx1864",
    journal = "Mon. Not. Roy. Astron. Soc.",
    volume = "472",
    number = "1",
    pages = "141--165",
    year = "2017"
}

@ARTICLE{2005ApJ...623.1026H,
       author = {{Hynes}, R.~I.},
        title = "{The Optical and Ultraviolet Spectral Energy Distributions of Short-Period Black Hole X-Ray Transients in Outburst}",
      journal = {\apj},
         year = 2005,
        month = apr,
       volume = {623},
       number = {2},
        pages = {1026-1043},
          doi = {10.1086/428445},
archivePrefix = {arXiv},
       eprint = {astro-ph/0412531},
 primaryClass = {astro-ph},
       adsurl = {https://ui.adsabs.harvard.edu/abs/2005ApJ...623.1026H}
}

@ARTICLE{2023AstBu..78....1B,
       author = {{Barsukova}, E.~A. and {Burenkov}, A.~N. and {Goranskij}, V.~P. and {Zharikov}, S.~V. and {Iliev}, L. and {Manset}, N. and {Metlova}, N.~V. and {Miroshnichenko}, A.~S. and {Moiseeva}, A.~V. and {Nedialkov}, P.~L. and {Semenko}, E.~A. and {Stoyanov}, K. and {Yakunin}, I.~A.},
        title = "{B[e] Star CI Camelopardalis in the Optical Range}",
      journal = {Astrophysical Bulletin},
         year = 2023,
        month = mar,
       volume = {78},
       number = {1},
        pages = {1-24},
          doi = {10.1134/S1990341323010029},
archivePrefix = {arXiv},
       eprint = {2412.12305},
 primaryClass = {astro-ph.SR},
       adsurl = {https://ui.adsabs.harvard.edu/abs/2023AstBu..78....1B}
}

@article{Angelini:2003sj,
    author = "Angelini, Lorella and White, Nicholas E.",
    title = "{An XMM-Newton observation of 4U1755-33 in quiescence: Evidence for a fossil x-ray jet}",
    eprint = "astro-ph/0302315",
    archivePrefix = "arXiv",
    doi = "10.1086/374682",
    journal = "Astrophys. J. Lett.",
    volume = "586",
    pages = "L71--L76",
    year = "2003"
}

@article{Kalemci:2018tkp,
    author = "Kalemci, Emrah and Maccarone, Thomas J. and Tomsick, John A.",
    title = "{A Dust-scattering Halo of 4U 1630{\textendash}47 Observed with Chandra and Swift: New Constraints on the Source Distance}",
    eprint = "1804.02909",
    archivePrefix = "arXiv",
    primaryClass = "astro-ph.HE",
    doi = "10.3847/1538-4357/aabcd3",
    journal = "Astrophys. J.",
    volume = "859",
    number = "2",
    pages = "88",
    year = "2018"
}

@article{Parker:2016ltr,
    author = "Parker, M. L. and others",
    title = "{NuSTAR and Swift observations of the very high state in GX 339-4: Weighing the black hole with X-rays}",
    eprint = "1603.03777",
    archivePrefix = "arXiv",
    primaryClass = "astro-ph.HE",
    doi = "10.3847/2041-8205/821/1/L6",
    journal = "Astrophys. J. Lett.",
    volume = "821",
    number = "1",
    pages = "L6",
    year = "2016"
}

@article{reid2023distances,
  title={On the distances to the X-ray binaries Cygnus X-3 and GRS 1915+ 105},
  author={Reid, MJ and Miller-Jones, JCA},
  journal={The Astrophysical Journal},
  volume={959},
  number={2},
  pages={85},
  year={2023},
  publisher={IOP Publishing}
}

@unpublished{Inokuchi:2025mch,
    author = "Inokuchi, Jin and Kawabata, Koji S. and Uemura, Makoto and Hiraga, Hiroyuki",
    title = "{Polarimetric Study of GRS 1915+105: Estimation of Interstellar Polarization Component}",
    eprint = "2501.02835",
    archivePrefix = "arXiv",
    primaryClass = "astro-ph.IM",
    month = "1",
    year = "2025"
}

@article{MacDonald:2014gpa,
    author = "MacDonald, Rachel K. D. and Bailyn, Charles D. and Buxton, Michelle and Cantrell, Andrew G. and Chatterjee, Ritaban and Kennedy-Shaffer, Ross and Orosz, Jerome A. and Markwardt, Craig B. and Swank, Jean H.",
    title = "{The Black Hole Binary V4641 Sagitarii: Activity in Quiescence and Improved Mass Determinations}",
    eprint = "1401.4190",
    archivePrefix = "arXiv",
    primaryClass = "astro-ph.SR",
    doi = "10.1088/0004-637X/784/1/2",
    journal = "Astrophys. J.",
    volume = "784",
    pages = "2",
    year = "2014"
}

@ARTICLE{2018ApJ...863..103S,
       author = {{Su}, Yang and {Zhou}, Xin and {Yang}, Ji and {Chen}, Yang and {Chen}, Xuepeng and {Zhang}, Shaobo},
        title = "{The Large-scale Interstellar Medium of SS 433/W50 Revisited}",
      journal = {\apj},
         year = 2018,
        month = aug,
       volume = {863},
       number = {1},
          eid = {103},
        pages = {103},
          doi = {10.3847/1538-4357/aad04e},
archivePrefix = {arXiv},
       eprint = {1807.03737},
 primaryClass = {astro-ph.GA},
       adsurl = {https://ui.adsabs.harvard.edu/abs/2018ApJ...863..103S}
}

@unpublished{AL-Zetoun:2025fxg,
    author = "AL-Zetoun, Ala'a",
    title = "{Propagation of Galactic cosmic rays: the influence of anisotropic diffusion}",
    eprint = "2506.13314",
    archivePrefix = "arXiv",
    primaryClass = "astro-ph.HE",
    month = "6",
    year = "2025"
}

@article{Jansson:2012pc,
    author = "Jansson, Ronnie and Farrar, Glennys R.",
    title = "{A New Model of the Galactic Magnetic Field}",
    eprint = "1204.3662",
    archivePrefix = "arXiv",
    primaryClass = "astro-ph.GA",
    doi = "10.1088/0004-637X/757/1/14",
    journal = "Astrophys. J.",
    volume = "757",
    pages = "14",
    year = "2012"
}

@ARTICLE{Green2015,
       author = {{Green}, D.~A.},
        title = "{Constraints on the distribution of supernova remnants with Galactocentric radius}",
      journal = {\mnras},
         year = 2015,
        month = dec,
       volume = {454},
       number = {2},
        pages = {1517-1524},
          doi = {10.1093/mnras/stv1885},
archivePrefix = {arXiv},
       eprint = {1508.02931},
 primaryClass = {astro-ph.HE},
       adsurl = {https://ui.adsabs.harvard.edu/abs/2015MNRAS.454.1517G}
}

@ARTICLE{2022SciBu..67.2162D,
       author = {{Dampe Collaboration}},
        title = "{Detection of spectral hardenings in cosmic-ray boron-to-carbon and boron-to-oxygen flux ratios with DAMPE}",
      journal = {Science Bulletin},
         year = 2022,
        month = nov,
       volume = {67},
       number = {21},
        pages = {2162-2166},
          doi = {10.1016/j.scib.2022.10.002},
archivePrefix = {arXiv},
       eprint = {2210.08833},
 primaryClass = {astro-ph.HE},
       adsurl = {https://ui.adsabs.harvard.edu/abs/2022SciBu..67.2162D}
}

@ARTICLE{2019ApJ...877...76K,
       author = {{Kleimann}, Jens and {Schorlepp}, Timo and {Merten}, Lukas and {Becker Tjus}, Julia},
        title = "{Solenoidal Improvements for the JF12 Galactic Magnetic Field Model}",
      journal = {\apj},
         year = 2019,
        month = jun,
       volume = {877},
       number = {2},
          eid = {76},
        pages = {76},
          doi = {10.3847/1538-4357/ab1913},
archivePrefix = {arXiv},
       eprint = {1809.07528},
 primaryClass = {astro-ph.GA},
       adsurl = {https://ui.adsabs.harvard.edu/abs/2019ApJ...877...76K}
}

@unpublished{Kaci:2025gyb,
    author = "Kaci, Samy and Giacinti, Gwenael and Aharonian, Felix and Wang, Jie-Shuang",
    title = "{Microquasars as the major contributors to Galactic cosmic rays around the ''knee''}",
    eprint = "2510.01369",
    archivePrefix = "arXiv",
    primaryClass = "astro-ph.HE",
    month = "10",
    year = "2025"
}

@unpublished{Fang:2026ydz,
    author = "Fang, Ke and Halzen, Francis",
    title = "{The Cosmic-ray Knee as a Local Signature of Nearby PeVatrons}",
    eprint = "2601.05435",
    archivePrefix = "arXiv",
    primaryClass = "astro-ph.HE",
    month = "1",
    year = "2026"
}

@unpublished{LHAASO:2025ysm,
    author = "Cao, Zhen and others",
    collaboration = "LHAASO",
    title = "{Cygnus X-3: A variable petaelectronvolt gamma-ray source}",
    eprint = "2512.16638",
    archivePrefix = "arXiv",
    primaryClass = "astro-ph.HE",
    month = "12",
    year = "2025",
    note = "arXiv:2512.16638"
}

@article{Merten:2017mgk,
    author = "Merten, Lukas and Becker Tjus, Julia and Fichtner, Horst and Eichmann, Bj and Sigl, G.",
    collaboration = "CRPropa",
    title = "{CRPropa 3.1{\textemdash}a low energy extension based on stochastic differential equations}",
    eprint = "1704.07484",
    archivePrefix = "arXiv",
    primaryClass = "astro-ph.IM",
    doi = "10.1088/1475-7516/2017/06/046",
    journal = "JCAP",
    volume = "06",
    pages = "046",
    year = "2017"
}

@article{Neronov:2024ycp,
    author = "Neronov, Andrii and Oikonomou, Foteini and Semikoz, Dmitri",
    title = "{Multimessenger signature of cosmic rays from the microquasar V4641 Sgr propagating along a Galactic magnetic field line}",
    eprint = "2410.17608",
    archivePrefix = "arXiv",
    primaryClass = "astro-ph.HE",
    doi = "10.1103/PhysRevD.111.103025",
    journal = "Phys. Rev. D",
    volume = "111",
    number = "10",
    pages = "103025",
    year = "2025"
}

@article{Fender:1998bb,
    author = "Fender, Robert P. and Pooley, Guy G.",
    title = "{Infrared synchrotron oscillations in grs 1915+105}",
    eprint = "astro-ph/9806073",
    archivePrefix = "arXiv",
    doi = "10.1046/j.1365-8711.1998.01921.x",
    journal = "Mon. Not. Roy. Astron. Soc.",
    volume = "300",
    pages = "573",
    year = "1998"
}

@ARTICLE{2005Natur.436..819G,
       author = {{Gallo}, Elena and {Fender}, Rob and {Kaiser}, Christian and {Russell}, David and {Morganti}, Raffaella and {Oosterloo}, Tom and {Heinz}, Sebastian},
        title = "{A dark jet dominates the power output of the stellar black hole Cygnus X-1}",
      journal = {\nat},
         year = 2005,
        month = aug,
       volume = {436},
       number = {7052},
        pages = {819-821},
          doi = {10.1038/nature03879},
archivePrefix = {arXiv},
       eprint = {astro-ph/0508228},
 primaryClass = {astro-ph},
       adsurl = {https://ui.adsabs.harvard.edu/abs/2005Natur.436..819G}
}

@ARTICLE{2011A&A...528A..89B,
       author = {{Bosch-Ramon}, V. and {Perucho}, M. and {Bordas}, P.},
        title = "{The termination region of high-mass microquasar jets}",
      journal = {\aap},
         year = 2011,
        month = apr,
       volume = {528},
          eid = {A89},
        pages = {A89},
          doi = {10.1051/0004-6361/201016364},
archivePrefix = {arXiv},
       eprint = {1101.5049},
 primaryClass = {astro-ph.HE},
       adsurl = {https://ui.adsabs.harvard.edu/abs/2011A&A...528A..89B}
}

@article{Abaroa:2025ege,
    author = "Abaroa, Leandro and Romero, Gustavo E. and Bosch-Ramon, Valent{\'\i}",
    title = "{Microquasar remnants as hidden PeVatrons}",
    eprint = "2512.07781",
    archivePrefix = "arXiv",
    primaryClass = "astro-ph.HE",
    doi = "10.1051/0004-6361/202557202",
    journal = "Astron. Astrophys.",
    volume = "705",
    pages = "L4",
    year = "2026"
}

@unpublished{Yue:2026gdt,
    author = "Yue, Hua and Nie, Lin and Guo, Yi-Qing and Hu, Hong-Bo",
    title = "{Joint constraint on the propagation origin of the cosmic-ray spectral knee from energy spectrum and anisotropy observations}",
    eprint = "2601.17851",
    archivePrefix = "arXiv",
    primaryClass = "astro-ph.HE",
    month = "1",
    year = "2026"
}

@unpublished{LHAASO:2025mlf,
    author = "Cao, Zhen and others",
    collaboration = "LHAASO",
    title = "{Precise Measurement of Cosmic Ray Light and Helium Spectra above 0.1 Peta-electron-Volt}",
    eprint = "2511.05013",
    archivePrefix = "arXiv",
    primaryClass = "astro-ph.HE",
    month = "11",
    year = "2025"
}

@article{Giacinti:2017dgt,
    author = "Giacinti, G. and Kachelriess, M. and Semikoz, D. V.",
    title = "{Reconciling cosmic ray diffusion with Galactic magnetic field models}",
    eprint = "1710.08205",
    archivePrefix = "arXiv",
    primaryClass = "astro-ph.HE",
    doi = "10.1088/1475-7516/2018/07/051",
    journal = "JCAP",
    volume = "07",
    pages = "051",
    year = "2018"
}

@article{Ohira:2024qtr,
    author = "Ohira, Yutaka",
    title = "{Very-high-energy gamma rays from cosmic rays escaping from Galactic black hole binaries}",
    eprint = "2410.22976",
    archivePrefix = "arXiv",
    primaryClass = "astro-ph.HE",
    doi = "10.1093/mnras/staf1176",
    journal = "Mon. Not. Roy. Astron. Soc.",
    volume = "2434",
    pages = "2439",
    year = "2025"
}

@article{Ptuskin2012,
  author  = {Ptuskin, V. and Zirakashvili, V.},
  title   = {Maximum Energies of Accelerated Particles in Supernova Remnants},
  journal = {Astronomy \& Astrophysics},
  year    = {2012},
  volume  = {543},
  pages   = {A47},
  doi     = {10.1051/0004-6361/201118643},
  url     = {https://doi.org/10.1051/0004-6361/201118643},
  note    = {Key later work modeling maximum particle energies in SNRs, including the role of ion-neutral damping and magnetic field amplification.},
}

@article{LHAASO:2021cbz,
    author = "Cao, Zhen and others",
    collaboration = "LHAASO*{\textdagger}, LHAASO",
    title = "{Peta{\textendash}electron volt gamma-ray emission from the Crab Nebula}",
    eprint = "2111.06545",
    archivePrefix = "arXiv",
    primaryClass = "astro-ph.HE",
    doi = "10.1126/science.abg5137",
    journal = "Science",
    volume = "373",
    number = "6553",
    pages = "425--430",
    year = "2021"
}

@article{HESS:2016pst,
    author = "Abramowski, A. and others",
    collaboration = "H.E.S.S.",
    title = "{Acceleration of petaelectronvolt protons in the Galactic Centre}",
    eprint = "1603.07730",
    archivePrefix = "arXiv",
    primaryClass = "astro-ph.HE",
    doi = "10.1038/nature17147",
    journal = "Nature",
    volume = "531",
    pages = "476",
    year = "2016"
}

@article{Unger:2023lob,
    author = "Unger, Michael and Farrar, Glennys R.",
    title = "{The Coherent Magnetic Field of the Milky Way}",
    eprint = "2311.12120",
    archivePrefix = "arXiv",
    primaryClass = "astro-ph.GA",
    doi = "10.3847/1538-4357/ad4a54",
    journal = "Astrophys. J.",
    volume = "970",
    number = "1",
    pages = "95",
    year = "2024"
}

@article{Podsiadlowski:2002ww,
    author = "Podsiadlowski, Philipp and Rappaport, Saul and Han, Zhanwen",
    title = "{On the formation and evolution of black hole binaries}",
    eprint = "astro-ph/0207153",
    archivePrefix = "arXiv",
    doi = "10.1046/j.1365-8711.2003.06464.x",
    journal = "Mon. Not. Roy. Astron. Soc.",
    volume = "341",
    pages = "385",
    year = "2003"
}

@article{Genolini:2019ewc,
    author = "G{\'e}nolini, Y. and others",
    title = "{Cosmic-ray transport from AMS-02 boron to carbon ratio data: Benchmark models and interpretation}",
    eprint = "1904.08917",
    archivePrefix = "arXiv",
    primaryClass = "astro-ph.HE",
    doi = "10.1103/PhysRevD.99.123028",
    journal = "Phys. Rev. D",
    volume = "99",
    number = "12",
    pages = "123028",
    year = "2019"
}

@ARTICLE{2026arXiv260706360O,
      author = {{Olivera-Nieto}, Laura and {Cowie}, Fraser J. and {Markoff}, Sera and {Fender}, Rob and {Crook-Mansour}, Justine},
       title = "{Extreme particle acceleration in X-ray binaries is linked to their jets}",
     journal = {arXiv e-prints},
        year = 2026,
       month = jul,
         eid = {arXiv:2607.06360},
       pages = {arXiv:2607.06360},
         doi = {10.48550/arXiv.2607.06360},
archivePrefix = {arXiv},
      eprint = {2607.06360},
primaryClass = {astro-ph.HE},
      adsurl = {https://ui.adsabs.harvard.edu/abs/2026arXiv260706360O}
}

@article{Johannesson:2016rlh,
    author = "J{\'o}hannesson, G. and others",
    title = "{Bayesian analysis of cosmic-ray propagation: evidence against homogeneous diffusion}",
    eprint = "1602.02243",
    archivePrefix = "arXiv",
    primaryClass = "astro-ph.HE",
    reportNumber = "IPPP-16-10",
    doi = "10.3847/0004-637X/824/1/16",
    journal = "Astrophys. J.",
    volume = "824",
    number = "1",
    pages = "16",
    year = "2016"
}

@article{Maurin:2022gfm,
    author = "Maurin, D. and Ferronato Bueno, E. and Derome, L.",
    title = "{A simple determination of the halo size from 10Be/9Be data}",
    eprint = "2203.07265",
    archivePrefix = "arXiv",
    primaryClass = "astro-ph.HE",
    doi = "10.1051/0004-6361/202243546",
    journal = "Astron. Astrophys.",
    volume = "667",
    pages = "A25",
    year = "2022"
}

@article{AL-Zetoun:2020pfg,
    author = "L-Zetoun, A. A. and Achterberg, A.",
    title = "{The contribution of nearby supernova remnants on the cosmic ray flux at Earth}",
    eprint = "2001.07602",
    archivePrefix = "arXiv",
    primaryClass = "astro-ph.GA",
    doi = "10.1093/mnras/staa171",
    journal = "Mon. Not. Roy. Astron. Soc.",
    volume = "493",
    number = "2",
    pages = "1960--1981",
    year = "2020"
}

@article{Aharonian:2026tzf,
    author = "Aharonian, Felix and Zhang, Bing Theodore",
    title = "{A Minimal Interpretation of the Galactic Cosmic-Ray Proton and Helium Spectra from GeV to PeV Energies}",
    eprint = "2602.08223",
    archivePrefix = "arXiv",
    primaryClass = "astro-ph.HE",
    doi = "10.3847/1538-4357/ae81a7",
    journal = "Astrophys. J.",
    volume = "1006",
    number = "2",
    pages = "211",
    year = "2026"
}

@article{Case:1998qg,
    author = "Case, Gary L. and Bhattacharya, Dipen",
    title = "{A new sigma-d relation and its application to the galactic supernova remnant distribution}",
    eprint = "astro-ph/9807162",
    archivePrefix = "arXiv",
    reportNumber = "UCRHEA-0798-01",
    doi = "10.1086/306089",
    journal = "Astrophys. J.",
    volume = "504",
    pages = "761",
    year = "1998"
}

@ARTICLE{Yu:2026vco,
       author = {{Yu}, Shiqi and {Zhang}, Bing Theodore},
        title = "{Galactic Microquasar and Supernova Remnants Imprinting on Diffuse Neutrino and Gamma-Ray Sky}",
      journal = {arXiv e-prints},
         year = 2026,
        month = aug,
          eid = {arXiv:2608.22374},
        pages = {arXiv:2608.22374},
          doi = {10.48550/arXiv.2608.22374},
archivePrefix = {arXiv},
       eprint = {2608.22374},
 primaryClass = {astro-ph.HE},
       adsurl = {https://ui.adsabs.harvard.edu/abs/2026arXiv260822374Y}
}

\end{document}